\documentclass[twocolumn]{jpsj2} 
\newcommand{\Romannum}[1]{\uppercase\expandafter{\romannumeral#1}}
\usepackage{bm}
\usepackage{mathrsfs}
\usepackage{exscale}
\usepackage{color}
\title{Microscopic Coexistence of Ferromagnetism and Superconductivity in Single-Crystal UCoGe}
\author{
Tetsuya \textsc{Ohta}$^{1}$,
Taisuke \textsc{Hattori}$^{1}$\thanks{E-mail address: t.hattori@scphys.kyoto-u.ac.jp}, 
Kenji \textsc{Ishida}$^{1,2}$\thanks{E-mail address: kishida@scphys.kyoto-u.ac.jp},
Yusuke \textsc{Nakai}$^{1,2}$,
Eisuke \textsc{Osaki}$^{3}$,
Kazuhiko \textsc{Deguchi}$^{3}$,
Noriaki K. \textsc{Sato}$^{3}$,
and
Isamu \textsc{Satoh}$^{4}$
}

\inst{
$^{1}$Department of Physics, Graduate School of Science, Kyoto University, Kyoto 606-8502, Japan.\\
$^{2}$Transformative Research Project on Iron Pnictides (TRIP), Japan Science and Technology Agency (JST), Tokyo 102-0075,  Japan.\\
$^{3}$Department of Physics, Graduate School of Science, Nagoya University, Nagoya 464-8602, Japan.\\
$^{4}$Institute for Materials Research, Tohoku University, Sendai 980-8577 Japan.
}

\abst{
Unambiguous evidence for the microscopic coexistence of ferromagnetism and superconductivity in UCoGe ($T_{\rm Curie} \sim 2.5$ K and $T_{\rm SC}$ $\sim$ 0.6 K) is reported from $^{59}$Co nuclear quadrupole resonance (NQR). The $^{59}$Co-NQR signal below 1 K indicates ferromagnetism throughout the sample volume, while nuclear spin-lattice relaxation rate $1/T_1$ in the ferromagnetic (FM) phase decreases below $T_{\rm SC}$ due to the opening of the superconducting(SC) gap. The SC state was found to be inhomogeneous, suggestive of a self-induced vortex state, potentially realizable in a FM superconductor. In addition, the $^{59}$Co-NQR spectrum around $T_{\rm Curie}$ show that the FM transition in UCoGe possesses a first-order character, which is consistent with the theoretical prediction that the low-temperature FM transition in itinerant magnets is generically of first-order. 
}

\kword{ferromagnetic superconductor, U-based heavy-fermion, UCoGe, Nuclear Quadrupole Resonance} 

\begin {document}
\maketitle

After the discovery of superconductivity in UGe$_2$ under pressure\cite{SaxenaNature00}, the coexistence of superconductivity and ferromagnetism becomes one of the major topics in condensed-matter physics. This is because ferromagnetism and spin-singlet superconductivity are thought to be mutually exclusive\cite{GinzburgJETP57,FayPRB80}. In the presence of a large splitting between the majority and minority spin Fermi surfaces, as in a ferromagnetic (FM) state, more-exotic spin-triplet superconductivity is allowed, in which parallel spins pair within each spin Fermi surface. While FM superconductors such as UIr\cite{AkazawaJPhys04} and URhGe\cite{AokiNature01} has recently been demonstrated to occur experimentally, proof that the same charge carriers participate simultaneously in both phenomena has remained elusive. 

In 2007, new ambient-pressure ferromagnetic (FM) superconductor UCoGe was discovered by Huy {\it et al}\cite{HuyPRL07}. UCoGe is a weak ferromagnet with $T_{\rm Curie} = 3$ K and the ordered moments $\mu_s = 0.03 \mu_{\rm B}$, and shows superconductivity at the transition temperature $T_{\rm SC} = 0.8$ K,\cite{HuyPRL07} highest within FM superconductors. In order to investigate the correlation between ferromagnetism and superconductivity, nuclear quadrupole resonance (NQR) measurements are ideally suited, since they provide microscopic information about the electronic and magnetic properties without applying external fields. In a magnetically ordered state, the NQR signal splits or shifts due to internal fields at the nuclear site, and the nuclear spin-lattice relaxation rate $1/T_1$ provides site-selective information about the density of states at the Fermi level and thus about the superconducting (SC) gap structure. UCoGe is a FM superconductor suitable for NQR measurements, since it contains an NQR-active element of $^{59}$Co.

In the previous letter, we reported $^{59}$Co-NQR studies in a polycrystalline UCoGe with $T_{\rm Curie}$ = 2.5 K and the SC onset temperature $T_{\rm SC}^{\rm onset}$ = 0.7 K\cite{OhtaJPSJ08}. We found inhomogeneous ferromagnetism below $T_{\rm Curie}$ in the polycrystalline sample, from the observation of the FM and nonmagnetic NQR spectra at lowest temperature. In addition, the SC anomaly was observed in $1/T_1$ measured in both NQR spectra, suggesting that superconductivity exists both in the FM and paramagnetic (PM) state. However, to investigate intrinsic nature of UCoGe, $^{59}$Co-NQR studies on a high-quality single crystal are highly desired.      

Another interest in UCoGe is its FM transition. The thermal FM transition at $T_{\rm Curie}$ in zero field is an well-known example of second-order phase transition. However, Belitz {\it et al.} theoretically suggested that the sufficiently low-temperature phase transition in itinerant ferromagnetism is generically of the first order\cite{BelitzPRL99}. Experimental results supporting this have been reported in a few ferromagnets\cite{UhlarzPRL04,PfleidererNature04}. In order to investigate the nature of the quantum FM transition occurring at $T$ = 0, UCoGe is one of the best ferromagnets, since $T_{\rm Curie}$ of UCoGe is as low as 2.5 K.

For the above purposes, we have performed $^{59}$Co-NQR measurements on two samples: one is the polycrystalline sample reported in literature\cite{OhtaJPSJ08} and the other is a 55 mg single crystal with $1.65 \times 1.65 \times 1.89$ mm$^3$ dimension, grown by the Czochralski method in a tetra-arc furnace. The sample preparation and characterization will be published elsewhere\cite{SatoSample}. The residual resistivity ratio along the $a$ axis for the single-crystal samples is 20. The FM transition temperature $T_{\rm Curie}$ of the single-crystal sample was evaluated to be $2.45 \pm 0.1$ K from the Arrot plots shown in Fig.~1 (a), where magnetization ($M$) was measured in fields ($H)$ parallel to the $c$ axis. Ising-type anisotropic behavior was observed with the easy axis along $c$. The magnitude of $m_0$ was determined to be 0.07 $\mu_B$/Co by extraporating at $T$ = 0. The $ac$ susceptibility, shown in Fig.1 (b), exhibits onset and midpoint SC transition temperatures, 0.70 and 0.57 K respectively in the single-crystal sample. The FM and SC properties of our single crystal are in good agreement with those obtained by Huy {\it et al}\cite{HuyPRL08} and D. Aoki {\it et al.}\cite{AokiUCoGe}. 
\begin{figure}[htbp]
\begin{center}
\includegraphics[clip=,width=0.75\columnwidth]{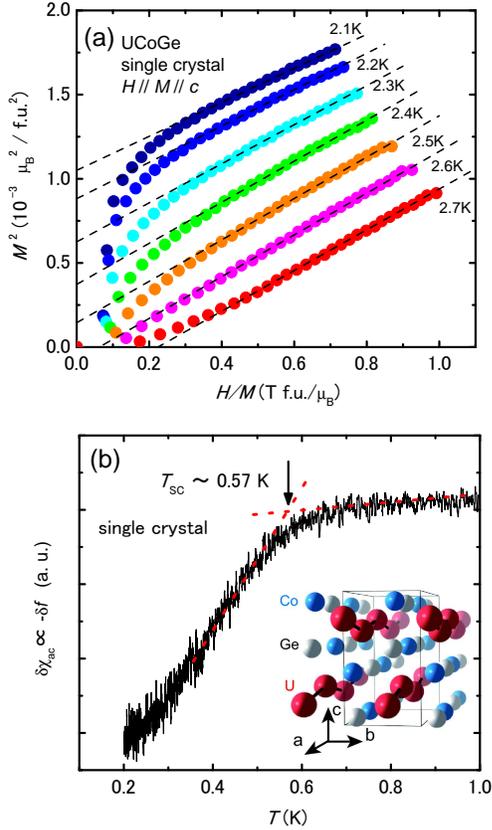}
\end{center}
\caption{(Color online)(a) Arrot plots of magnetization isotherms measured in fields $H \parallel c$ at temperatures from 2.1 to 2.7 K. (b) Temperature dependence of the ac susceptibility $\chi_{ac}$ in zero field. Inset: Crystal structure of UCoGe (orthorhombic TiNiSi structure), drawn in Vesta\cite{VESTA}. }
\end{figure}

\begin{figure}[htbp]
\begin{center}
\includegraphics[clip=,width=0.75\columnwidth]{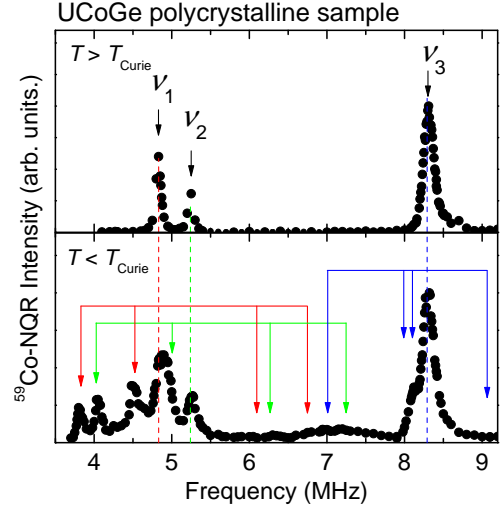}
\end{center}
\caption{(color online) $^{59}$Co-NQR spectra in the polycrystalline sample obtained (a) at 4.2 K in the PM state and (b) at 95mK in the FM state. The arrows in the FM state are the calculated resonance frequencies (see text)  }
\end{figure}

Figure 2 (a) shows the $^{59}$Co-NQR ($I$ = 7/2) spectrum observed at 4.2 K in the PM state of the polycrystalline sample, exhibiting three peaks from the $m = \pm 1/2 \leftrightarrow \pm 3/2 (\nu_1)$, $\pm 3/2 \leftrightarrow \pm 5/2 (\nu_2)$, and $\pm 5/2 \leftrightarrow \pm 7/2 (\nu_3)$ transitions. From the resonance peaks, the NQR frequency of the principal-axis component of the electric field gradient (EFG), $|\nu_{zz}|$, is evaluated to be 2.85 MHz, and the asymmetric parameter $\eta$ defined as $(\nu_{xx} - \nu_{yy})/\nu_{zz}$ is 0.52. The direction of the principal axis of the EFG is calculated to be tilted by 10$^{\circ}$ from the $a$ axis in the $ac$ plane. The EFG principal axis is parallel to the zig-zag chain consisting of the U atoms\cite{Harima}. In the FM state of the polycrystalline sample, the complicated $^{59}$Co-NQR spectrum shown in Fig. 2 (b) was observed at 95 mK. We found signals around 4 and 7 MHz, which could not be observed previously\cite{OhtaJPSJ08}.  
When the nuclear spin experiences an internal field $\mbox{\boldmath$H$}_{\rm int}$, the Zeeman term $H_{\rm Z} = -\gamma_{\rm n} \hbar \mbox{\boldmath$I$} \cdot \mbox{\boldmath$H$}_{\rm int}$ = $-\gamma_{\rm n}\hbar H_{\rm int} (I_z\cos \theta + I_x \sin \theta )$ is added in the electric-quadrupole interaction, where $\gamma_{\rm n}$ is the Co nuclear gyromagnetic ratio, the $z$ component of $I$ is along the EFG principal axis, $\theta$ is the angle between the EFG principal axis and $\mbox{\boldmath$H$}_{\rm int}$, which is in the $ac$ plane for the crystallographical symmetry. The NQR spectrum shown in Fig. 2 (b) is consistently understood by the superposition of the PM NQR spectra without $\mbox{\boldmath$H$}_{\rm int}$ and the FM NQR spectra with ($H_{\rm int}, \theta$) = (910 Oe, 80$^{\circ}$).  The internal-field direction from the ordered U moments is tilted by 80$^{\circ}$ from the the EFG principal axis in the $ac$ plane, which makes the U ordered moments consistent with the $c$-axis Ising anisotropy observed.  

\begin{figure}[htbp]
\begin{center}
\includegraphics[clip=,width=0.75\columnwidth]{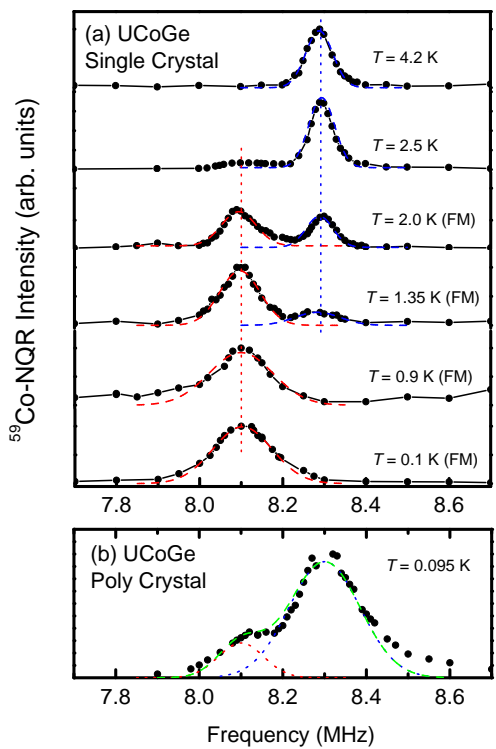}
\end{center}
\caption{ (a) Temperature dependence of the NQR spectrum from the $\pm 5/2 \Leftrightarrow \pm 7/2$ transitions ($\nu_3$) in the single-crystal sample. The blue (red) broken lines represent Gaussian fits to the PM (FM) peaks; the solid lines are guides to the eye. (b) NQR spectrum of the $\nu_3$ transition in the polycrystalline sample at 95 mK. }
\end{figure}
In the single-crystal UCoGe, although three NQR signals were observed in the PM state, we could not observe signals associated with $\nu_1$ and $\nu_2$ in the FM state. The undetection of $\nu_1$ and $\nu_2$ signals would be due to a weak signal/noise ratio and/or short $T_2$ effect. In the following, focusing on the NQR signal around 8.3 MHz ($\nu_3$), we discuss the character of a FM transition in UCoGe and differences between the polycrystalline and single-crystal samples. Figure 3 (a) shows the temperature variation of the spectrum in the single-crystal sample. The full width at half maximum (FWHM) of the 8.3 MHz peak is 70 kHz at 4.2 K, half that observed in the polycrystalline sample, indicating the single-crystal sample has a more homogeneous EFG and higher quality. With decreasing temperature, the intensity of the 8.3 MHz NQR signal arising from the PM region decreases below $\sim 3.7$ K while the 8.1 MHz signal originating from the FM region appears below 2.7 K. The two NQR signals coexist between 1 and 2.7 K, but the PM signal disappears below 0.9 K. This is in contrast with the temperature variation in the polycrystalline sample, where the PM $^{59}$Co-NQR signal remained even at $T = 95$ mK shown in Fig.~3 (b). The narrow temperature range of the coexistence in the single-crystal sample is likely indicative of better sample quality. We stress that the single-crystal UCoGe is in the homogeneous FM state throughout the sample below 1 K from the absence of the PM signal.      

Notably, the temperature dependence of the NQR spectrum in UCoGe through $T_{\rm Curie}$ is quite different from that observed in a second-order transition. For example, in the antiferromagnet CeRhIn$_5$ (N\`eel temperature $T_N \sim 3.8$ K), the $\pm 5/2 \leftrightarrow \pm 7/2$ transition in $^{115}$In-NQR ($I = 9/2$) shifts continuously to lower frequencies below $T_N$, indicating the continuous development of the ordered moment\cite{MitoPRB01}. It is also noteworthy that the resonance frequency of the FM phase is nearly temperature independent, indicating that the internal field remains constant below $T_{\rm Curie}$. These are in agreement with the theoretical prediction that low-temperature itinerant FM transitions are generically of a first order\cite{BelitzPRL99}. 

\begin{figure}[htbp]
\begin{center}
\includegraphics[clip=,width=0.75\columnwidth]{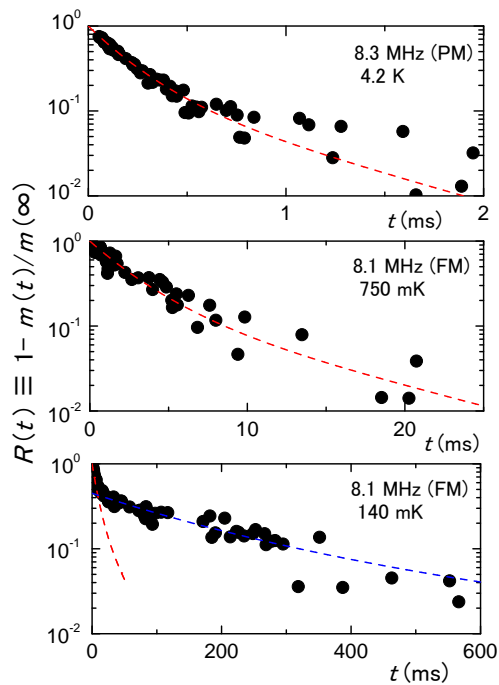}
\end{center}
\caption{(Color online) Recovery curves $R(t)$ of the nuclear magnetization $m(t)$ at time $t$ after a saturation pulse with the fits used to evaluate $1/T_1$\cite{Narath,ChepinJPhys91}. Two relaxation components were clearly observed below $T_{\rm SC}$ in the single-crystal sample.}
\end{figure}
\begin{figure}[htbp]
\begin{center}
\includegraphics[clip=,width=0.75\columnwidth]{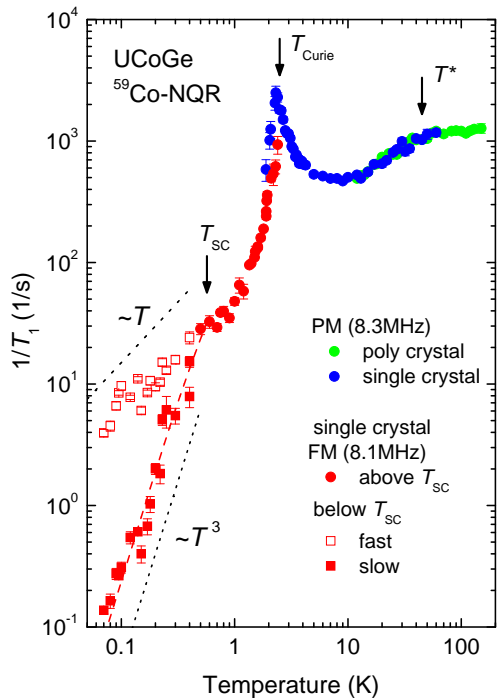}
\end{center}
\caption{(Color online) Temperature dependence of $^{59}$Co $1/T_1$ in the single-crystal sample, along with $1/T_1$ in the polycrystal between 10 and 150 K. $1/T_1$ was measured at the PM (8.3 MHz) frequency above 2.3 K, shown by closed blue (green) circles for the single-crystal (polycrystalline) sample.  Below 2.3 K, $1/T_1$ was measured at the FM (8.1 MHz) frequency. Two $1/T_1$ components were observed in the SC state, the faster (slower) component denoted by red open (closed) squares; the red broken curve below $T_{\rm SC}$ represents the temperature dependence calculated assuming line-node gap and $\Delta_0/k_BT_{\rm SC}$ = 2.3.} 
\end{figure}

The temperature dependence of $1/T_1$ at 8.3 and 8.1 MHz provides information about spin dynamics related to the U moments in the normal state and SC properties below $T_{\rm SC}$. Figures 4 show the recovery curves $R(t) = 1 - m(t) / m(\infty)$ of the nuclear magnetization $m(t)$ measured on the single-crystal sample at 4.2, 0.75, and 0.14 K. Here, $m(t)$ is the nuclear magnetization at a time $t$ after a saturation pulse. The $R(t)$ data measured at the PM 8.3 MHz signal at 4.2 K can be fit consistently by the theoretical function for $I = 7/2$ with a single $T_1$ component\cite{Narath,ChepinJPhys91}. Below 2.3 K, $1/T_1$ was measured at the FM 8.1 MHz $^{59}$Co-NQR peak and is still described by a single component down to $T_{\rm SC}$. Below $T_{\rm SC}$, slower relaxation component was observed in $R(t)$ and this tendency is more pronounced as temperature is lowered. $R(t)$ in the SC state possesses nearly equal amount of the fast and slow components, and thus the fast (slow) relaxation rate 1/$T_1$ was determined by fitting the recovery curve of $0.5 < R(t) < 1 (0.01 < R(t) < 0.5)$ region, as shown in Fig.~4 (c)     

Figure 5 shows the temperature dependence of $1/T_1$ in the single-crystal UCoGe down to 70 mK, together with the polycrystalline results between 10 and 150 K.  The $1/T_1$s in the single-crystal and polycrystalline samples agree well between 10 and 60 K, remaining nearly constant down to $T^* \sim 40$ K and gradually decreasing below $T^*$. Since the magnetic susceptibility $\chi$ deviates from the Curie-Weiss behavior and the electrical resistivity along the $c$ axis shows metallic behavior below about $T^*$\cite{SatoSample}, $T^*$ is regarded as the characteristic temperature below which the U 5$f$ electrons become itinerant with relatively heavy electron mass. Below 10 K, $1/T_1$ increases to a remarkable peak at $T_{\rm Curie} \sim$ 2.5 K due to the critical slowing down of U moments. Note that $\chi\equiv\chi(q=0)$ also diverges, indicative of FM ordering in zero field. The strong divergence of $1/T_1$ implies that the FM transition in UCoGe would be weakly first order, since the critical fluctuations are not usually observed in a first-order transition. 

In the SC state, the fast component of $1/T_1$ at the FM signal is roughly proportional to $T$, consistent with Korringa behavior characteristic of metals, indicating that it originates from non-superconducting regions. In contrast, the slow component at the FM signal decreases rapidly below $T_{\rm SC}$, roughly as $T^3$, suggestive of line nodes in a SC gap. The red broken line in Fig.~5 shows a fit using the line-node model $\Delta(\theta) = \Delta_0\cos\theta$ with $\Delta_0 = 2.3k_BT_{\rm SC}$. The detection of the SC gap via the FM signal makes this strong unambiguous evidence for microscopic coexistence. Although this is the second piece of evidence for microscopic coexistence of itinerant ferromagnetism and superconductivity after $^{73}$Ge-NQR in UGe$_2$ under pressure\cite{KotegawaJPSJ05}, the present result on UCoGe is more unambiguous, since the $^{59}$Co-NQR signals from the PM and FM states are well separated (see Fig.~3 (a)) and the NQR measurements were performed on a single crystal at ambient pressure.

The results below $T_{\rm SC}$ provide some new insight on the nature of the superconductivity in UCoGe. From the relaxation in the FM signal, nearly half of the sample's volume remains non-superconducting even at 70 mK, while the sample is in a homogeneous FM state below $T_{\rm Curie}$. Similar two-relaxation rate behavior was previously observed in the polycrystalline sample\cite{OhtaJPSJ08}. Since the temperature where the second component emerges coincides with $T_{\rm SC}$ for both single-crystal and polycrystalline samples, the two-relaxation rate behavior would be intrinsic. One possibility is a nonunitary SC state, which is realized in the superfluid $^{3}$He-$A_1$ under an external magnetic field\cite{LeggettRMP75}. In such a SC state, although only up-spin pair is formed, the SC state is spatially homogeneous. The inhomogeneous SC state in UCoGe appears to be incompatible with the homogeneous nonunitary SC state. Alternative interpretation of this behavior is a self-induced vortex (SIV) state\cite{TachikiSSC80}. When superconductivity occurs in a FM state, it has been suggested that a SIV state in which vortices are generated spontaneously can be stable for $H_{c1} < 4\pi M < H_{c2}$. There, the regions near vortex cores, where the SC gap is largely suppressed would give rise to the fast component of $1/T_1$. Our single-crystal results suggest that the SC gap is inhomogeneous in a real space as in the polycrystalline sample\cite{OhtaJPSJ08}, which appears to be consistent with the SIV state. The recent observation of a slight increase in the muon decay rate below $T_{\rm SC}$ might be related to the SIV state, since the SIV would produce a distribution of internal fields at the implanted muon site\cite{VisserPRL09}. The SIV state has been discussed theoretically, but has never been identified experimentally. UCoGe is a promising candidate in which the SIV state may be realized.

In conclusion, $^{59}$Co-NQR measurements on UCoGe show the first-order character of the FM transition, and the unambiguous evidence for the microscopic coexistence of ferromagnetism and superconductivity. The coexistence originates from the same U-$5f$ electrons, ruling out real-space phase separation. Although ferromagnetism exists homogeneously throughout the sample, UCoGe's superconductivity would be intrinsically inhomogeneous, which might be interpreted in terms of a SIV state; further work, particularly low-temperature STM/STS measurements in zero field, will be important to detect the vortices produced by FM moments in the SC state.

The authors thank D. C. Peets, S. Yonezawa, H. Takatsu, S. Kittaka, and Y. Maeno for experimental support and valuable discussions, and H. Harima, H. Ikeda, S. Fujimoto, A. de Visser, D. Aoki, and J. Flouquet for fruitful discussions. This work was partially supported by Kyoto Univ. LTM center, the "Heavy Electrons" Grant-in-Aid for Scientific Research on Innovative Areas  (No. 20102006) from MEXT of Japan, a Grant-in-Aid for the Global COE Program ``The Next Generation of Physics, Spun from Universality and Emergence'' from MEXT of Japan, a grant-in-aid for Scientific Research from JSPS and KAKENHI (S) (No. 20224015) from JSPS.

\end{document}